\documentclass[twocolumn,showpacs,preprintnumbers]{revtex4}
\usepackage{graphicx}
\usepackage{dcolumn}
\usepackage{bm}
\usepackage{epsfig}
\usepackage{amsmath}
\usepackage{mathtools}
\usepackage{subfigure}
\usepackage{epstopdf}
\usepackage{hyperref}
\usepackage{mathrsfs,bm,amsthm,amsfonts,xspace,float,bookmark}
\usepackage{textcomp}

\begin{document}

\title{Dark state cooling of a trapped ion using microwave coupling}
\author{Yong Lu$^{1,2}$}
\author{Jian-Qi Zhang$^{3}$}
\author{Jin-Ming Cui$^{1,2}$}
\author{Dong-Yang Cao$^{1,2}$}
\author{Shuo Zhang$^{2,4,5}$}
\email{zhang31415926@gmail.com}
\author{Yun-Feng Huang$^{1,2}$}
\email{hyf@ustc.edu.cn}
\author{Chuan-Feng Li$^{1,2}$}
\email{cfli@ustc.edu.cn}
\author{Guang-Can Guo$^{1,2}$}
\address{$^1$Key Laboratory of Quantum Information, University of Science and Technology of China, Hefei 230026, P. R. China
\\$^2$Synergetic Innovation Center of Quantum Information and Quantum Physics, University of Science and Technology of China, Hefei 230026, P. R. China
\\ $^3$State Key Laboratory of Magnetic Resonance and Atomic and Molecular Physics,
Wuhan Institute of Physics and Mathematics, Chinese Academy of Sciences, Wuhan 430071, China
\\ $^4$Zhengzhou Information Science and Technology Institute, Zhengzhou, 450004, China
\\ $^5$ Science and Technology on Information Assurance Laboratory, Beijing 100071, China
}

\begin{abstract}
We propose a new dark-state cooling method of trapped ion systems in the Lamb-Dicke limit. With application of microwave
dressing the ion, we can obtain two electromagnetically induced transparency structures. The heating effects caused by the carrier and the blue sideband transition vanish due to the EIT effects and the final mean phonon numbers can be much less than the recoil limit. Our scheme is robust to fluctuations of microwave power and laser intensities which provides a broad cooling bandwidth to cool motional modes of a linear ion chain. Moreover, it is more suitable to cool four-level ions on a large-scale ion chip.
\pacs{37.10.Rs, 42.50.-p}
\end{abstract}
\maketitle

\section{INTRODUCTION}

\label{sec1}

Cooling a single trapped ion to the ground state has been a fundamental subject in quantum metrology, quantum simulation and quantum computation \cite{1,2,3,4}. Sideband cooling is the first method to reach the vibrational ground state of a single trapped-ion and has been implemented experimentally \cite{5,6,7}. Two requirements should be satisfied for sideband cooling: (1) the ion should be localized to dimensions much smaller than the optical wavelength (Lambe-Dicke limit, LDL), and (2) the absorption sidebands should be well-resolved (strong confinement) ,i.e., the vibrational frequency should be much larger than the natural linewidth $\Gamma$ of the optical transition for laser cooling. Although the sideband cooling can cool the ion close to the ground state, the final mean phonon number is still higher than the recoil limit due to the off-resonant carrier and blue sideband transitions.

To prevent the off-resonant carrier transition, quantum coherent effects have been exploited in several schemes, e.g., the cavity-assisted cooling scheme \cite{8,9,10}, and the dark-state cooling schemes \cite{11,12,13,14,15,16,17,18}. The electromagnetically induced transparency (EIT) cooling is the first dark state cooling proposed by G.Morigi et al \cite{11} which has been a standard experimental scheme to cool trapped ions \cite{16,19}, and cooling a nanomechanical resonator to its ground state has been proposed based on EIT cooling \cite{20,21}. EIT cooling uses a three-level trapped-ion system coupled by two lasers with the same blue detuning, a Fano-like absorption spectrum with a zero at the carrier transition and a peak at the first-order red sideband transition. Therefore the EIT cooling scheme can achieve a lower temperature than sideband cooling. Moreover no strong confinement is required due to the three level nature of the system.

Further improvements in approaches to reach the ground state are achieved with the double-dark state concepts by eliminating all heating transitions. According to the EIT cooling scheme \cite{11}, the double-EIT cooling \cite{12} results in a double-dark state for the two EIT structures, thus eliminating both the carrier and the blue sideband transition. An alternative approach that combines the methods of EIT and Stark shift cooling \cite{13} is the robust and fast cooling scheme \cite{14}, which cancels both off-resonant excitations due to the combination of EIT and Stark shift Hamiltonian contributions. Dark-state cooling using standing waves \cite{17} localizes the ion at the node of the laser to eliminate the carrier transition, and the blue sideband heating is prevented by the EIT effect \cite{22}. Ground-state cooling by quantum interference pathways \cite{17} and fast cooling through magnetic gradients \cite{18} can also cancel all the dominating heating effects. In the above schemes, all the leading-order heating effects are suppressed, and the whole system can be cooled to a steady state, namely, the double dark state.
\begin{equation}
\left| \Psi  \right\rangle=|a_0\rangle |0\rangle  + \beta |a_1\rangle \left| 1 \right\rangle  + O\left( {{\eta ^2}} \right).
\label{eq1}
\end{equation}
where $\left|{a_0}\right\rangle$, $\left|{a_1}\right\rangle$ are the atomic internal states, $\left|{0}\right\rangle$, $\left|{1}\right\rangle$ are the vibrational ground state and lowest excitation respectively, and $\beta$ is a coefficient of the order $O\left({{\eta}}\right)$. Thus, the final phonon number is zero in the leading order of the expansion in the Lamb-Dicke parameter $\eta$, i.e,
\begin{equation}
{n_{ss}}=0+O\left({\eta^2}\right).
\label{eq2}
\end{equation}

The scheme presented here also eliminates the carrier and blue sideband terms completely (up to the first-order in the Lambe-Dicke parameter), using an atomic four-level system, a combination of two laser beams and a microwave field. The application of a microwave field is used to dress two bare atomic ground states. By tuning the Rabi frequency of the microwave field to the condition $\Omega=-\nu/2$, two EIT structures show up to cancel both the heating transitions.

Our scheme has several advantages in experimental implementation: (1) similar to the double-EIT cooling \cite{10}, EIT cooling \cite{11,16} and sympathetic EIT cooling \cite{19}, our cooling scheme requires relatively small laser intensities and provides a broad cooling bandwidth such that multiple motional modes of a linear ion chain can be cooled simultaneously compared to previous implementations of conventional Raman sideband cooling. However, the scheme here uses a microwave instead of one of three laser beams in the double-EIT cooling, and microwave manipulation components are inherently simpler, more stable and scalable compared to lasers \cite{24,25,26,27,28}; (2) for many cooling setups \cite{14,17,18,23} based on a $\Lambda$-configuration, they requires additional repumping fields because the excitation state decays to more than two lower states while the repumping process can disturb corresponding double-dark states, and worsen cooling performance. Instead of repumping the system arbitrarily, we use a suitable microwave to improve cooling efficiency. Therefore with our scheme the unwanted additional decay pathways can become an advantage. In addition, unwanted upper levels can also affect the cooling result seriously \cite{16,19}, while it has a limited affect to cool the atoms with nuclear spins by using our cooling scheme.

The rest of this paper is organized as follows. In Sec. \uppercase\expandafter{\romannumeral2}, the cooling scheme and corresponding Hamiltonian are presented. Under the appropriately chosen microwave Rabi frequency and lasers' detuning, the system can stay in a double-dark steady state and all the main heating effects are cancelled. In Sec. \uppercase\expandafter{\romannumeral3}, the analytical formula for the cooling rate is obtained by the method of adiabatic elimination and the optimum condition for the cooling process is derived. Sec. \uppercase\expandafter{\romannumeral4} provides numerical simulations of the cooling dynamics and the robustness against the laser and microwave intensity fluctuation. Finally, a brief conclusion is given.

\section{Cooling scheme and Hamiltonian}

\label{sec2}
\begin{figure}[tbph]
\includegraphics[width=6cm]{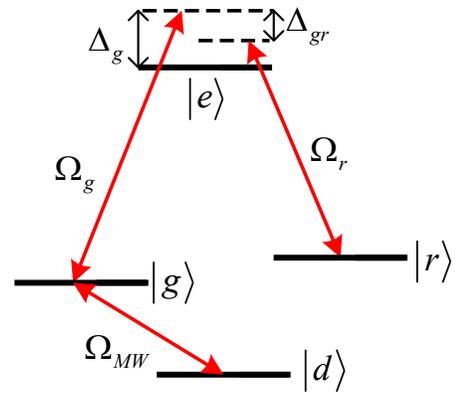}
\caption{(Color online) The four-level atom has an excited state $|e\rangle$, and three ground states $|g\rangle$,
$|r\rangle$ and $|d\rangle$. $|e\rangle\leftrightarrow|g\rangle$ and $|e\rangle\leftrightarrow|r\rangle$ are coupled by the
two laser beams while $|g\rangle\leftrightarrow|d\rangle$ is driven by a microwave field.}
\label{fig1}
\end{figure}

Our scheme is shown in Fig. 1. A single ion with mass M is trapped in a one-dimensional harmonic trap with the vibrational frequency $\nu$. $|n\rangle$ represents the $n$-th phonon state; $b^{\dag}$and $b$ are the phonon's creation and annihilation operators. The internal states of the ion consist of three nondissipation states, $|d\rangle$, $|r\rangle$ and $|g\rangle$, and one dissipative state, $|e\rangle$, and the corresponding energies are $\hbar\omega_0$, $\hbar\omega_r$, $\hbar\omega_g$ and $\hbar\omega_e$, respectively. The linewidth of state $|e\rangle$ is $2\gamma$. The transitions $|e\rangle\leftrightarrow|g\rangle$ and $|e\rangle\leftrightarrow|r\rangle$ are driven by two lasers with frequencies $\omega_{L_1}$ and $\omega_{L_2}$, wave vectors $k_{L_1}$ and $k_{L_2}$, and Rabi frequencies $\Omega_r$ and $\Omega_g$, respectively. States $\left|d\right\rangle$ and $\left|g\right\rangle$ are coupled to a resonant microwave field with frequency $\omega_{MW}$ and Rabi frequency $\Omega_{MW}$. The Hamiltonian of the whole system reads $\left(\hbar=0\right)$
\begin{equation}
\begin{array}{rcl}
H &= & \nu{b^{\dag} }b - {\Delta _g}\left| e \right\rangle \left\langle e \right| - \left( {{\Delta _g} - {\Delta _r}} \right)\left| r \right\rangle \left\langle r \right|  \\
&+& {\Omega _g}\left| e \right\rangle \left\langle g \right|{e^{i{k_{{L_1}}}x}} + {\Omega _r}\left| e \right\rangle \left\langle r \right|{e^{i{k_{{L_2}}}x}} \\
 &+& {\Omega _{MW}}\left| g \right\rangle \left\langle d \right| + H.c,
\end{array}
\label{eq3}
\end{equation}
where $\Delta_g$, $\Delta_r$ are the detunings of the two lasers, respectively , which are

\begin{equation}
\begin{array}{rcl}
\Delta _g &=& {\omega _{L_1}} - \left({{\omega _e} - {\omega _g}} \right), \\
\Delta _r &= &  {\omega _{L_2}} - \left( {{\omega _e} - {\omega _r}} \right),\\
\Delta_{gr}& =& \Delta_g-\Delta_r.
\end{array}
\label{eq004}
\end{equation}
The cooling process occurs in the Lamb-Dicke (LD) limit. Thus, the Hamiltonian (\ref{eq3}) can be expanded by the LD parameters around the trap center:
\begin{equation}
H_{LD}={H_{at}} + {H_m} + V,
\end{equation}
where
\begin{equation}
{H_m}=\nu{b^{\dag}}b
\end{equation}
is the free Hamiltonian for vibrational degrees of freedom (DOF),

\begin{equation}
\begin{array}{rcl}
H_{at}&= &-{\Delta _g}\left| e \right\rangle \left\langle e \right| - {\Delta _{gr}}\left| r \right\rangle \left\langle r \right| \\
&+& \left( {{\Omega _g}\left| e \right\rangle \left\langle g \right| + {\Omega _r}\left| e \right\rangle \left\langle r \right| + {\Omega _{MW}}\left| g \right\rangle \left\langle d \right| + H.c.} \right).
\end{array}
\label{eq7}
\end{equation}
represents the atomic internal DOF, and
\begin{equation}
V= i{\eta _1}{\Omega _g}\left| e \right\rangle \left\langle g \right|\left( b+b^{\dag} \right) + i{\eta _2}{\Omega _r}\left| e \right\rangle \left\langle r \right|\left(b +b^{\dag}\right) + H.c.
\end{equation}
describes the sideband transitions due to the two lasers. Here, the LD parameters are defined by
\begin{equation}
{\eta _1} = {k_{{L_1}}}\sqrt {\frac{1}{{2M\nu}}}, {\eta _2} = {k_{{L_2}}}\sqrt {\frac{1}{{2M\nu}}}.
\end{equation}

If $\Omega _{MW}=\Delta _{gr}$, we can rewrite the Hamiltonian of the atomic internal DOF (\ref{eq7}) in a new representation $\{\left|D\right\rangle,\left|B\right\rangle,\left|+\right\rangle,\left|e\right\rangle\}$
\begin{equation}
\begin{array}{rcl}
H_{at} &=& - \left( {{\Delta _g} - {\Delta _{gr}}} \right)\left| e \right\rangle \left\langle e \right| + 2{\Delta _{gr}}\left|  +  \right\rangle \left\langle  +  \right| \\
& + & \left( {{\Omega _+}\left|  +  \right\rangle \left\langle e \right| + {\Omega _B}\left| B \right\rangle \left\langle e \right| + H.c.} \right),
\end{array}
\label{eq10}
\end{equation}
where
\begin{equation}
\begin{array}{rcl}
{\Omega _B} &=& \frac{1}{\sqrt 2 }\sqrt {2\Omega _r^2 + \Omega _g^2},\\
{\Omega _ + }&=& \frac{1}{\sqrt 2 }\Omega _g,\\
\left|  +  \right\rangle & = & \frac{1}{\sqrt 2 }\left( {\left| g \right\rangle  + \left| d \right\rangle } \right), \\
\left| B \right\rangle & = & \frac{1}{{\sqrt {2\Omega _r^2 + \Omega _g^2} }}\left( {{\Omega _g}\left|  -  \right\rangle  + {\sqrt 2} {\Omega _r}\left| r \right\rangle } \right).
\end{array}
\label{eq11}
\end{equation}
It can be found that for the dark state $\left|D\right\rangle$, which is
\begin{equation}
\left| D \right\rangle = \frac{1}{{\sqrt {2\Omega _r^2 + \Omega _g^2} }}\left( {\sqrt 2 {\Omega _r}\left| -\right\rangle -{\Omega _g}\left| r \right\rangle }\right),
\end{equation}
where $\left|  -  \right\rangle=\frac{1}{{\sqrt 2 }}\left( {\left| g \right\rangle  - \left| 0 \right\rangle } \right)$, the state $\left|D\right\rangle$ decouples from the Hamiltonian (\ref{eq10}), which is the internal steady state, and the carrier transition vanishes due to the EIT effect. Then we can rewrite the sideband transition V with the representation $\{\left|D\right\rangle,\left|B\right\rangle,\left|+\right\rangle,\left|e\right\rangle\}$ as follows:,
\begin{equation}
\begin{array}{rcl}
V &=& i{\eta _D}{\Omega _D}\left| D \right\rangle \left\langle e \right|(b+b^{\dag})+ H.c.\\
&+& i{\eta _1}{\Omega _ + }\left|  +  \right\rangle \left\langle e \right|(b+b^{\dag}) + H.c.\\
&+& i{\eta _B}{\Omega _B}\left| B\right\rangle \left\langle e \right|(b+b^{\dag}) + H.c.
\end{array}
\end{equation}
with
\begin{equation}
{\eta _D }= {\eta _1} - {\eta _2}, \eta _B = \frac{{\Omega _g^2{\eta _1} + 2\Omega _r^2{\eta _2}}}{{2\Omega _r^2 + \Omega _g^2}}, {\Omega _D} = \frac{{{\Omega _g}{\Omega _r}}}{{\sqrt {2\Omega _r^2 + \Omega _g^2} }}.
\label{eq14}
\end{equation}

\begin{figure}[tbph]
\includegraphics[width=8cm]{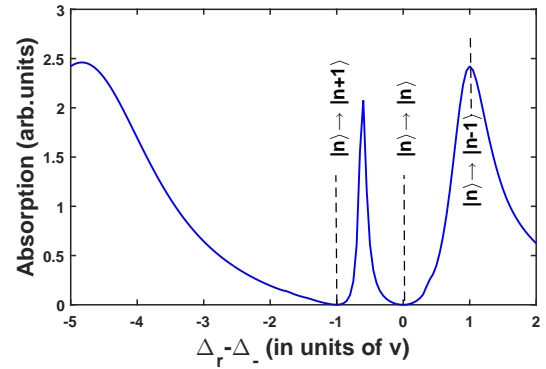}
\caption{(Color online) Absorption of cooling laser. As there are two EIT structures, the absorption vanishes at two values for the detuning  $\Delta_r$ of the cooling laser. Dashed lines mark the positions of these zeros and the absorption maximum. For appropriately chosen parameters, both the carrier transition and the blue-sideband heating are completely cancelled, while the red-sideband cooling is maximized. Here, the parameters are $\Omega_g = \gamma/\protect\sqrt{2}, \Omega_r =\gamma/20, \Delta_- = \gamma$ which represents the detuning for the transition $|-\rangle\leftrightarrow |e\rangle$. The other parameters are chosen to satisfy Eq.(\protect\ref{eq15}) and Eq.(\protect\ref{eq32}).}
\label{fig2}
\end{figure}

To understand the cooling dynamics, it is convenient to transfer the Hamiltonian to a dressed representation$\{|+\rangle, |-\rangle, |r\rangle, |e\rangle\}$ , which is
\begin{equation}
\begin{array}{rcl}
H_{at} &=&\Delta_r|r\rangle\langle r|+\Delta_-|-\rangle\langle -|+\Delta_{+}|+\rangle\langle+| \\
& + & \frac{\Omega _g}{\sqrt 2}(|+\rangle\langle e| + | e\rangle\langle  +|)+\frac{\Omega _g}{\sqrt 2}(|-\rangle\langle e| + | e\rangle\langle -|)\\
& + & \Omega_r(|r\rangle\langle e| + | e\rangle\langle r|).
\end{array}
\end{equation}
with
\begin{equation}
\begin{array}{rcl}
\Delta_-&=&\Delta_{r},\\
\Delta_+&=&\Delta_r+2\Delta_g,
\end{array}
\end{equation}
where $\Delta_r$, $\Delta_+$, $\Delta_-$ are corresponding detunings for transitions $|r\rangle\leftrightarrow|e\rangle$, $|+\rangle\leftrightarrow|e\rangle$, $|-\rangle\leftrightarrow|e\rangle$, respectively. The transitions $\left|-\right\rangle \leftrightarrow \left|e\right\rangle$ and $\left|r\right\rangle \leftrightarrow \left|e\right\rangle$ have the same detunings to produce the dark state $|D\rangle$. Hence the carrier transition is eliminated, and the transitions $\left|D\right\rangle \left|n\right\rangle\leftrightarrow \left|e\right\rangle\left|n+1\right\rangle $ and $\left|+\right\rangle \left|n+1\right\rangle\leftrightarrow \left|e\right\rangle\left|n+1\right\rangle $ remain. If detuning $\Delta_{gr}$ satisfies
\begin{equation}
\Delta _{gr}= -\frac{\nu}{2},
\label{eq15}
\end{equation}
that is $\Delta_+=\Delta_r-\nu$, then  EIT will arise between $\left|D\right\rangle\left|n\right\rangle\leftrightarrow\left|e\right\rangle\left|n+1\right\rangle $ and $\left|+\right\rangle \left|n+1\right\rangle\leftrightarrow\left|e\right\rangle\left|n+1\right\rangle $, and the blue sideband heating is coherently canceled while only the cooling transition is left. As expected, Fig.\ref{fig2} shows a typical scattering spectrum of the cooling laser (the cooling transition $|r\rangle\leftrightarrow |e\rangle$). The spectrum can be understood as consisting of two Fano-like structures, which is characteristic of EIT, and is controlled by the strong coupling laser field. Under the conditions Eq.\ref{eq15} and Eq.\ref{eq32} all the heating transitions vanish, while the maximal red-sideband cooling is obtained. Therefore, the cooling cycle will start from the $n$th vibrational steady $|D\rangle|n\rangle$, scatter over the excited state $|e\rangle|n-1\rangle$ via the red sideband transition, and dissipate back to the lower vibrational steady state $|D\rangle|n-1\rangle$, and the final double-dark state is
\begin{equation}
\left| \psi  \right\rangle  = {\Omega _ + }\left| D \right\rangle \left| 0 \right\rangle  - i{\eta _D}{\Omega _D}\left|  +  \right\rangle \left| 1 \right\rangle  + O\left( {{\eta ^2}} \right),
\end{equation}
which satisfies $H_{LD}|\psi\rangle=0$. The final temperature can be evaluated by the mean phonon number of the double-dark state, yielding
\begin{equation}
{n_{ss}} = \frac{{\eta _D^2\Omega _D^2}}{{\eta _D^2\Omega _D^2 + \Omega _ + ^2}} + O\left( {{\eta ^4}} \right).
\end{equation}
According to (\ref{eq11}) and (\ref{eq14}),
\begin{equation}
{n_{ss}} = \frac{{2\eta _D^2\Omega _g^2\Omega _r^2}}{{\Omega _g^4 + 2\left( {\eta _D^2 + 1} \right)\Omega _g^2\Omega _r^2}} + O\left( {{\eta ^4}} \right).
\label{eq18}
\end{equation}
If we choose ${\Omega _g} \gg {\Omega _r}$, the mean phonon number $n_{ss}$ will be much lower than the phonon recoil limit $O\left(\eta^2\right)$.

\section{ANALYTICAL TREATMENT}

\label{sec3}

The master equation is useful for analytically characterizing the cooling dynamics. Both the coherent dynamics and the dissipative nature of the excited level can be accounted for as follows:

\begin{equation}
\frac{d}{dt}\rho=\mathcal{L}\rho= - i[H,\rho ] + {\mathcal{L}_{\gamma}}\rho,
\end{equation}
where the Liouvilian $\mathcal{L}_{\gamma}$ is for three dissipative channels, and
\begin{equation}
\begin{split}
   \mathcal{L}_\gamma \rho & = \sum_{j = g, r, d} {\gamma _j} \bigg( {2\int_{ - 1}^1 {dsW( s ){e^{i{k_j}xs}}| j \rangle \langle e |\rho | e \rangle \langle j |{e^{ - i{k_j}xs}}} }  \\
    & - | e \rangle \langle e |\rho  - \rho | e \rangle \langle e | \bigg),
\end{split}
\end{equation}
where $2\gamma_j$ is the spontanous decay rate from $\left|e\right\rangle$ to $\left|j\right\rangle\left(j=g,r,d\right)$, $W\left(s\right)$ is the angular distribution of a spontaneous emission for a dipole transition, and $k_j=\left(\omega_e-\omega_j\right\rangle\// c$.

In view of the small magnitude of $\eta$, the coupling between the internal and external DOF (i.e., the laser cooling) is slow compared with the time scale of the internal atomic dynamics in the LDL. This situation allows adiabatic elimination of the internal DOF, and thus we can obtain the rate equation of the motional state
\begin{equation}
\begin{array}{rcl}
\dfrac{d}{dt} {p_{n}} & = & (n + 1)A_-p_{n+1}+nA_+p_{n-1}\\
& - &\left[(n+1)A_++nA_-\right]p_n,
\end{array}
\end{equation}
where $ p_n=\left\langle n \right|\rho \left| n \right\rangle$ is the population of the nth vibrational state and $ A_{\pm}$ are the cooling and heating transition rates.

Then the rate equation for the mean phonon number $\left\langle n\right\rangle$ is
\begin{equation}
\frac{d}{dt}\left\langle n\right\rangle = -\left(A_{-}-A_{+}\right)\left\langle n\right\rangle + A_{+}.
\end{equation}

If $A_{+}<A_{-}$, the mean phonon number varies with time, as follows:
\begin{equation}
{\langle n\rangle} = \left({\langle n\rangle}_{0}-{\langle n\rangle}_{ss}\right)e^{-Wt}+{\langle n\rangle}_{ss},
\label{eq23}
\end{equation}
where ${\langle n\rangle}_{0}$ is the initial mean phonon number; ${\langle n\rangle}_{ss} $ is the final mean phonon number; and $W$ is the rate of the cooling process:
\begin{align}
{\langle n\rangle}_{ss}  & = \frac{A_{+}}{A_{-}-A_{+}}+O(\eta^2), \\
W & = A_{-}-A_{+}.
\end{align}

To obtain the cooling transition rates $A_{\pm}$, we can define $\Gamma_{n\rightarrow n\pm 1}$ as the cooling and heat scattering amplitudes:
\begin{equation}
\Gamma_{n\rightarrow n\pm 1} = 2\gamma{\left|\langle n \pm 1 |\langle D |W_\gamma G(E_n)V|D\rangle|n\rangle\right|}^2,
\end{equation}
where $\Gamma_{n\rightarrow n-1}$ and $\Gamma_{n\rightarrow n+1}$ describe the cooling and heating scattering process from the $n$th-vibration steady state $|D\rangle|n\rangle$ to the $(n\pm 1)$th-vibration steady state $|D\rangle|n\pm 1\rangle$ via the sideband transition $V$. $E_n$ is the energy of state $|D\rangle|n\rangle$; the operator $W_\gamma=|D\rangle\langle e |$ describes the dissipation back to the internal dark state $|D\rangle$ from the excited state $|e\rangle$; and the function $G(z)$ is defined by
\begin{equation}
G(z)=\frac{1}{z-H_{\mathrm{eff}}},
\end{equation}
where the effective Hamiltonian $H_{\mathrm{eff}}=H_{LD}-i\gamma |e\rangle\langle e|$.

Calculating Eq.\ref{eq15}, we can get
\begin{equation}
\Gamma_{n\rightarrow n\pm 1}=2\gamma {\Omega_D ^2}{\eta ^2}(n+\delta_{\pm}){\nu^2}\frac{(\mp \nu-2\Delta_{gr})^2}{|f(\mp \nu)|^2},
\end{equation}
where $\delta_{+}=1,\delta_{-}=0$, and
\begin{equation}
\begin{split}
f(x) & =x^3+(\Delta_g-\Delta_{gr})x^2 \\
&+({\Omega_B}^2+{\Omega_+}^2+2\Delta_g\Delta_{gr}+2\Delta_{gr}^2)x +2{\Omega_B}^2\Delta_{gr}\\
&+i\gamma(x-2\Delta_{gr})x.
\end{split}
\end{equation}
Because $\Gamma_{n\rightarrow (n\pm 1)}=(n \pm \delta_{\pm})A_{\pm}$, the following holds:
\begin{equation}
A_{\pm}= 2\gamma {\Omega_D ^2}{\eta ^2}{\nu^2}\frac{(\mp\nu-2\Delta_{gr})^2}{f(\mp\nu)}.
\end{equation}

As expected, $A_+$ vanishes at $\Delta_{gr}=-\frac{\nu}2$, which means the carrier and the blue sideband heating are inhibited,and the final mean phonon number $n_{ss}$ will be zero in the leading order of the expansion in the LD parameter $\eta$. Then the corresponding cooling transition rate is
\begin{equation}
A_{n\rightarrow n-1} = 2\gamma {\Omega_D ^2}{\eta ^2}\frac{4\nu^2}{{\left(3\nu^2+2\nu\Delta_g-2\Omega_B^2-\Omega_+^2\right)^2+4{\gamma}^2\nu^2}}.
\end{equation}
We can obtain the maximal cooling rate at
\begin{equation}
3\nu^2+2\nu\Delta_g-2\Omega_B^2-\Omega_+^2=0.
\label{eq32}
\end{equation}
The maximum $W$ is
\begin{equation}
W = \frac{2\Omega_D ^2\eta ^2}{\gamma}.
\label{eq33}
\end{equation}

\section{NUMERICAL RESULTS AND DISCUSSION}

\label{sec4}

We have described the model and obtained the analytical result of the cooling process. In this section, we use the quantum toolbox \cite{29} for numerical simulation of the master equation Eq.\ref{eq3}. A $^{171}\textmd{Yb}^+$ ion is regarded as an excellent candidate for lager-scale quantum computation with qubits encoded in two hyperfine states \cite{26,27,30,31,32,33,34}. Our scheme can be experimentally implemented in the $^{171}\textmd{Yb}^+$ ion system, where the energy levels of $|r\rangle, |g\rangle, |d\rangle$ and $|e\rangle$ can be realized by choosing $|^{2}S_{1/2}, F=1, m_F=-1\rangle, |^{2}S_{1/2}, F=1, m_F=0\rangle, |^{2}S_{1/2}, F=1, m_F=1\rangle$ and $|^{2}P_{1/2}, F=0, m_F=0\rangle$, respectively.We can use two 369-nm laser beams to couple the transition $|^{2}S_{1/2}, F=1, m_F=-1\rangle\leftrightarrow |^{2}P_{1/2}, F=0, m_F=0\rangle$ and  $|^{2}S_{1/2}, F=1, m_F=0\rangle\leftrightarrow |^{2}P_{1/2}, F=0, m_F=0\rangle$, and use an acousto-optical modulator (AOM) driver to tune the detuning difference between these two transitions. To drive the transition  $|^{2}S_{1/2}, F=1, m_F=0\rangle\leftrightarrow |^{2}S_{1/2}, F=1, m_F=1\rangle$, a magnetic field can be used to split these two generate levels, and then the transition can be driven with a microwave field.The branching ratios out of the excitation state $|^{2}P_{1/2}, F=0, m_F=0\rangle$ to the three hyperfine states.
$|^{2}S_{1/2}, F=1, m_F=-1\rangle, |^{2}S_{1/2}, F=1, m_F=0\rangle $ and $|^{2}S_{1/2}, F=1, m_F=1\rangle$ are the same, that is $\gamma_g=\gamma_r=\gamma_d$.

\begin{figure}[t]
\includegraphics[width=8cm]{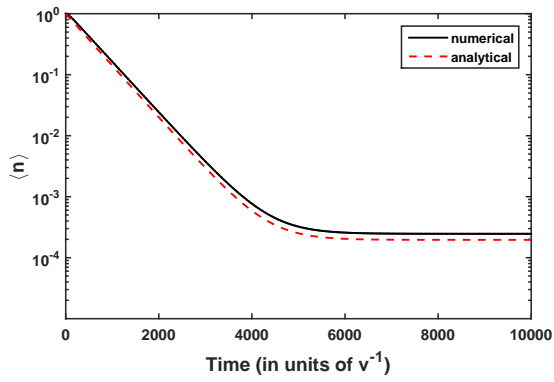}
\caption{(Color online) Numerical simulation of cooling a $^{171}\textmd{Yb}^{+}$. The parameters are $\eta_g=-\eta_r=0.05$, $2\gamma=20\nu$, $\gamma_g=\gamma_r=\gamma_d= 10\nu/3$, $\Omega_g=10\nu$, $\Omega_r=1\nu$, $\Omega_{MW}=\Delta_{gr}=-0.5\nu$, $\Delta_g=74.5\nu$. The analytical predictions are shown as dashed lines.The solid lines correspond to the numerical results. The final phonon number is much less than the recoil limit $O(10^{-2})$.}
\label{fig3}
\end{figure}
As shown in Fig. \ref{fig3}, the analytical prediction of the cooling dynamics in Eq.\ref{eq23} is consistent with the numerical simulation, and the final phonon number of our scheme can reach the order of $O(10^{-4})$. As $\eta_D=0.1$, the corresponding recoil energy is approximately $O(\eta _D^2)-O(10^{-2})$, and then the final temperature in our scheme can be much lower than the recoil energy for $\Omega_g\gg\Omega_r$.

\begin{figure}[tbph]
\includegraphics[width=8cm]{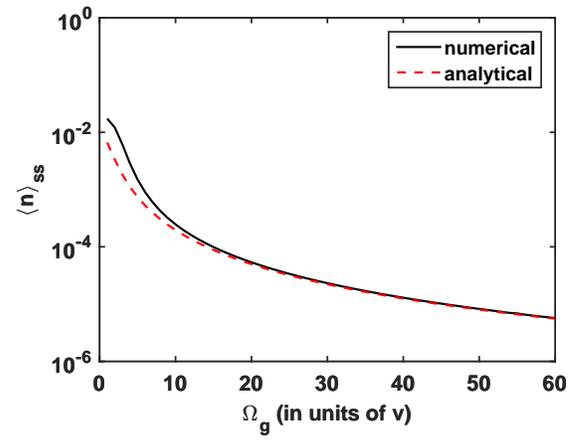}
\caption{(Color online) Numerical simulation of the final phonon numbers as a function of $\Omega_g$. The parameters are $\eta_g=-\eta_r=0.05$, $2\gamma=20\nu$, $\gamma_g=\gamma_r=\gamma_d= 10\nu/3$, $\Omega_r=1\nu$, $\Omega_{MW}=\Delta_{gr}=-0.5\nu$, $\Delta_g$ are chosen under the optimal condition Eq.(\protect\ref{eq32}). The dashed line is obtained by analytical prediction while the solid line corresponds to numerical results. The cooling result gets better with increasing $\Omega_g$.}
\label{fig4}
\end{figure}

\begin{figure}[tbph]
\includegraphics[width=8cm]{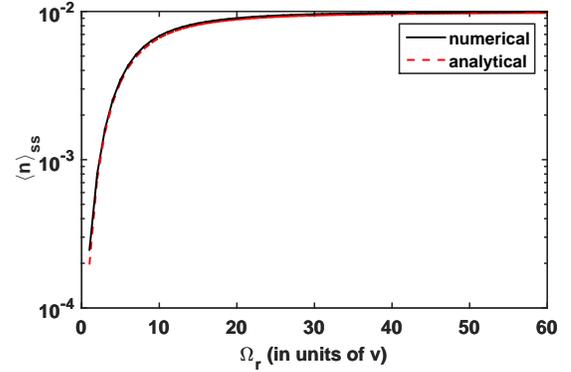}
\caption{(Color online) Numerical simulation of the final phonon numbers as a function of $\Omega_r$. The parameters are $\eta_g=-\eta_r=0.05$, $2\gamma=20\nu$, $\gamma_g=\gamma_r=\gamma_d= 10\nu/3$, $\Omega_g=10\nu$, $\Omega_{MW}=\Delta_{gr}=-0.5\nu$, $\Delta_g$ are chosen under the optimal condition Eq.(\protect\ref{eq32}).The dashed line is obtained with analytical prediction while the solid line corresponds to numerical results. The curves show the final phonon number decreases with increasing $\Omega_r$. }
\label{fig5}
\end{figure}

The dependence of the final cooling result on the Rabi frequencies $\Omega_g, \Omega_r$ is shown in Fig. \ref{fig4} and Fig. \ref{fig5}, respectively. In Fig. \ref{fig4}, with the fixed $\Omega_r$ and other parameters chosen under optimal conditions (\ref{eq15}) and (\ref{eq32}), the final mean phonon number decreases with increasing $\Omega_g$. In contrast, as the Rabi frequency $\Omega_g$ is fixed, the final cooling result worsens with increasing  $\Omega_r$, which is shown in Fig. \ref{fig5}. In the two figures, the numerical simulation agrees with the analytical prediction Eq.(\ref{eq18}). The final phonon number increases by increasing the strength of $\Omega_g$ while it decreases with increasing $\Omega_r$  because they affect the final double-dark according to Eq.\cite{18}.

\begin{figure}[tbph]
\includegraphics[width=8cm]{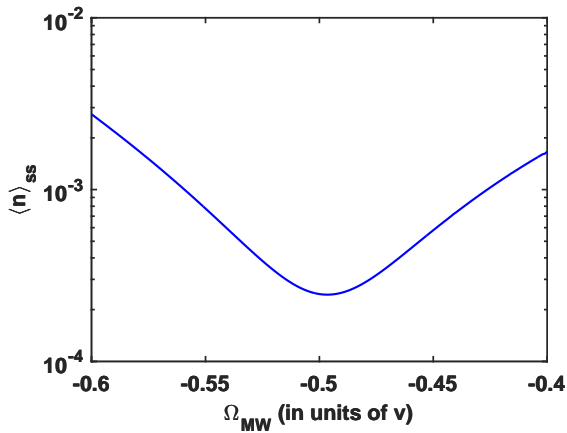}
\caption{(Color online) Numerical simulation of the final phonon numbers as a function of $\Omega_{MW}$. The parameters are $\eta_g=-\eta_r=0.05$, $2\gamma=20\nu$, $\gamma_g=\gamma_r=\gamma_d= 10\nu/3$, $\Omega_g=10\nu$, $\Omega_r=1\nu$, $\Delta_{gr}=-0.5\nu$, $\Delta_g=74.5\nu$. The corresponding optimal value of $\Omega_{MW}$ from the analytical prediction is at $\Omega_{MW0}=-0.5\nu$. The cooling result is robust against the fluctuation of the microwave driver's power $\Omega_{MW}.$}
\label{fig6}
\end{figure}
\begin{figure}[tbph]
\includegraphics[width=8cm]{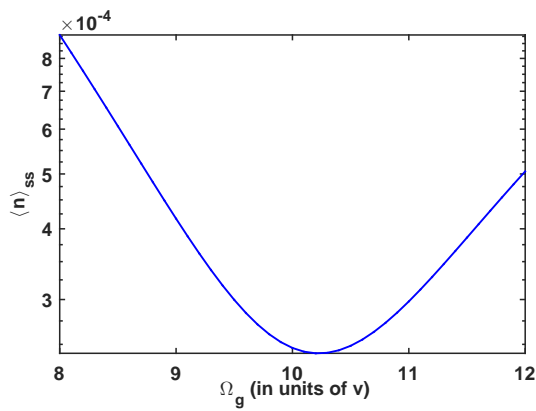}
\caption{(Color online) Numerical simulation of the final phonon numbers as a function of $\Omega_g$. The parameters are $\eta_g=-\eta_r=0.05$, $2\gamma=20\nu$, $\gamma_g=\gamma_r=\gamma_d= 10\nu/3$, $\Omega_r=1\nu$, $\Omega_{MW}=\Delta_{gr}=-0.5\nu$, $\Delta_g=74.5\nu$.The corresponding optimal value of $\Omega_g$ from the analytical prediction is at $\Omega_g=10\nu$. The cooling result is robust against the fluctuation of $\Omega_g$ that depends on the optical laser intensity.}
\label{fig7}
\end{figure}

Our cooling scheme is robust with respect to the fluctuation of the Rabi frequencies. As depicted in Fig. \ref{fig6} or Fig. \ref{fig7}, when the fluctuation of the Rabi frequency $\Omega_{MW}$ or $\Omega_g$  is approximately $20\%$, the final mean phonon number varies by the order of $10^{-3}$. Therefore, the cooling scheme is insensitive to the fluctuation of the microwave power and the laser optical intensity.

Our scheme is more suitable for the ions with hyperfine structures such as $^{171}\textmd{Yb}^{+}$ and $^{111}\textmd{Cd}^{+}$. For previous methods based on the three-level $\Lambda$-configuration, in experimental implementation with special ions such as $^{40}\textmd{Ca}^{+}$ and $ ^{24}\textmd{Mg}^{+}$, unwanted upper levels result in decreasing the cooling performance\cite{16,19}, whereas for $^{171}\textmd{Yb}^{+}$ and other parameters here, the unwanted hyperfine structure $|^{2}P_{1/2}, F=1\rangle$ which includes three hyperfine excitation states in radiant $2\gamma=19.7$ MHz has a natural 2.1 GHz gap with the level $|^{2}P_{1/2}, F=0, m_F=0\rangle$, and only $|^{2}P_{1/2}, F=1, m_F=-1\rangle$, $|^{2}P_{1/2}, F=0, m_F=1\rangle$ can lead to a weaker scattering rate $R\approx\frac{2}{3}\times(\frac{\Omega_g}{\Delta})^2\times2\gamma= 0.32$ KHz compared to the cooling rate $W=2$ KHz according to Eq.\cite{35}. Hence the cooling scheme here is efficient. Moreover, we only require one laser to implement Doppler cooling and our cooling scheme simultaneously instead of Raman sideband cooling.

Recently, large-scale quantum-computer architecture are more and more attractive \cite{24,25,26,27,28,30,31,32,33,34,35,36,37,38,39,40}, and scaling can potentially be achieved by storing ions in multizone arrays where information is moved in the processor by physically transporting the ions \cite{35,36} or teleporting \cite{37}. It is important to recool the qubit ion because ions can be heated by ambient noisy electric fields and/or during ion transport \cite{30}. Experimentally, ``refrigerant'' ions can be used to cool the qubit ions sympathetically \cite{4,19,35,36,38,39,40}. Microwave can be applied via waveguides (structures that can guide radiation) that are part of the chip on which the ion trap is integrated \cite{27}, and thus does not require alignment, while laser beams must be carefully aligned to interact with the trapped ions \cite{26}. Moreover, it is easier to generate and control microwave fields which are more robust to amplitude, phase, frequency and polarization noise compared to the use of lasers \cite{24,25}, and these parameters are difficult to control in integrated optics \cite{28}. Hence, compared to double-EIT cooling, our cooling scheme uses a microwave driver instead of one laser beam to ease the experimental condition for cooling ``refrigerant'' ions in multizone arrays, and make it more feasible to realize the double-dark state cooling on the ion chip firstly.

\section{CONCLUSION}

\label{sec5}

In conclusion, we have shown a new efficient double-dark state cooling scheme for a trapped ion system using the EIT effect combined with microwave Stark shift. Microwave coupling is used to dress the ion levels and obtain two EIT structures, providing a double dark state. In leading order of the Lamb-Dicke expansion, this suppresses all heating excitations. As a consequence, the final temperature in this scheme can be much lower than the recoil energy. It is robust against fluctuations of the microwave power and the laser intensities. Compared with other dark-state cooling methods, this scheme is better for the experimental realization of the cooling of some specific trapped ions.
\section*{ACKNOWLEDGMENTS}
We thank Y. Lin and Q. Yin for helpful comments on the manuscript. This work was supported by the National Natural Science Foundation of China (Nos. 61327901, 61490711, 11274289, 11325419, 11474268, 11304366, 11404319, 91421111, 61205108), the National Basic Research Program of China (No. 2011CB921200), the Strategic Priority Research Program (B) of the Chinese Academy of Sciences (Grant No. XDB01030300), the Fundamental Research Funds for the Central Universities (No. WK2470000011, No. WK2470000018), the China Postdoctoral Science Foundation (Grant Nos. 2013M531771 and 2014T70760), Foundation of Science and Technology on Information Assurance Laboratory (No. KJ-14-001);


\begin{thebibliography}{99}
\bibitem{1} Cirac, J. I. and P. Zoller, Phys.
Rev. Lett. \textbf{74}, 4091 (1995).

\bibitem{2} C. Monroe, D. M. Meekhof, B. E. King, W. M. Itano, and D. J. Wineland, Phys.
Rev. Lett. \textbf{75}, 4714 (1995)

\bibitem{3} Childs, A. M. and I. L. Chuang, Phys.
Rev. A. \textbf{63}, 012306 (2000).

\bibitem{4} P. O. Schmidt, T. Rosenband, C. Langer, W. M. Itano, J. C. Bergquist, and D. J. Wineland,
Science. \textbf{309}, 749-752 (2005)
\bibitem{5} F. Diedrich, J. C. Bergquist, W. M. Itano, and D. J. Wineland, Phys.
Rev. Lett. \textbf{62}, 403 (1989)
\bibitem{6} C. Monroe, D. M. Meekhof, B. E. King, S. R. Jefferts, W. M. Itano, D. J. Wineland, and P. Gould, Phys.
Rev. Lett. \textbf{75}, 4011 (1995)
\bibitem{7} C. Roos, T. Zeiger, H. Rohde, H. C. Nagerl, J. Eschner, D. Leibfried, F. Schmidt-Kaler, and R. Blatt, Phys.
Rev. Lett. \textbf{83}, 4713 (1999)
\bibitem{8} Stefano Zippilli and Giovanna Morigi, Phys.
Rev. A. \textbf{72}, 053408 (2005)
\bibitem{9} Marc Bienert and Giovanna Morigi, New.
J. Phys. \textbf{14}, 023002 (2012)
\bibitem{10} Shuo Zhang, Qian-Heng Duan, Chu Guo, Chun-Wang Wu, Wei Wu, and Ping-Xing Chen, Phys. Rev. A \textbf{89}, 013402 (2014)
\bibitem{11} Giovanna Morigi, J\"{u}rgen Eschner, and Christoph H. Keitel, Phys.
Rev. Lett. \textbf{85}, 4458 (2000)
\bibitem{12} J. Evers and C. H. Keitel, Europhys. Lett. \textbf{68}, 370 (2004)
\bibitem{13} A Retzker and M B Plenio, New.
J. Phys. \textbf{9}, 279 (2007)
\bibitem{14} J. Cerrillo, A. Retzker, and M. B. Plenio, Phys.
Rev. Lett. \textbf{104}, 043003 (2010)
\bibitem{15} Giovanna Morigi, Phys.
Rev. A. \textbf{67}, 033402 (2003)
\bibitem{16} C. F. Roos, D. Leibfried, A. Mundt, F. Schmidt-Kaler, J. Eschner, and R. Blatt , Phys.
Rev. Lett. \textbf{85}, 5547 (2000)
\bibitem{17} Shuo Zhang, Chun-Wang Wu, and Ping-Xing Chen, Phys.
Rev. A. \textbf{85}, 053420 (2012)
\bibitem{18} Shuo Zhang, Jian-Qi Zhang, Qian-Heng Duan, Chu Guo, Chun-Wang Wu, Wei Wu, and Ping-Xing Chen, Phys.
Rev. A. \textbf{90}, 043409 (2014)
\bibitem{19} Y. Lin, J. P. Gaebler, T. R. Tan, R. Bowler, J. D. Jost, D. Leibfried, and D. J. Wineland,
Phys. Rev. Lett. \textbf{110}, 153002 (2013)
\bibitem{20} Keyu Xia and Jorg Evers, Phys. Rev. Lett. \textbf{103}, 227203 (2009)
\bibitem{21} J. Q. Zhang, S Zhang, J. H. Zou, L. Chen, W. Yang, Y. Li,  and M. Feng, Optics Express \textbf{21}, 29695-29710 (2013)
\bibitem{22} Michael Fleischhauer, Atac Imamoglu, and Jonathan P. Marangos , Rev.
Mod. Phys. \textbf{77}, 633 (2005)
\bibitem{23} A. Albrecht, A. Retzker, C. Wunderlich and M. B. Plenio, New.
J. Phys. \textbf{13}, 033009 (2011)
\bibitem{24} C. Ospelkaus, U. Warring, Y. Colombe, K. R. Brown, J. M. Amini, D. Leibfried, D. J. Wineland, Nature. \textbf{476}, 181¨C184 (2011)
\bibitem{25} D. P. L. Aude Craik, N. M. Linke, T. P. Harty, C. J. Ballance, D. M. Lucas, A. M. Steane, D. T. C. Allcock, Applied Physics B.
\textbf{114}, 3-10 (2014)
\bibitem{26} W. K. Hensinger, Nature. \textbf{476}, 155 (2011)
\bibitem{27} M. D. Hughes, B. Lekitsch, J. A. Broersma and  W. K. Hensinger, Contemp. Phys. \textbf{52} 505¨C29 (2011)
\bibitem{28} G. R. Brady, A. R. Ellis, D. L. Moehring, D. Stick, C. Highstrete, K. M. Fortier, M. G. Blain, R. A. Haltli, A. A. Cruz-Cabrera, R. D. Briggs, J. R. Wendt, T. R. Carter, S. Samora, S. A. Kemme, Appl. Phys. B Lasers Opt. \textbf{103}(4), 801 (2011).
\bibitem{29} Sze.M.Tan, J.Opt. B. \textbf{4}, 424 (1999)
\bibitem{30} C. Monroe and J. Kim, Science \textbf{339}, 1164 (2013).
\bibitem{31} C. Monroe, R. Raussendorf, A. Ruthven, K. R. Brown, P. Maunz, L.-M. Duan, J. Kim, Phys. Rev. A \textbf{89}, 022317 (2014).
\bibitem{32} T. D. Ladd, F. Jelezko, R. Laflamme, Y. Nakamura, C. Monroe, and J. L. O¡¯Brien, Nature \textbf{464}, 45 (2010).
\bibitem{33} R. C. Sterling, H. Rattanasonti, S. Weidt, K. Lake, P. Srinivasan, S. C. Webster, M. Kraft ,and W. K. Hensinger, Nat.Commun.\textbf{5}, 3637, (2013)
\bibitem{34} Ch. Piltz, Th. Sriarunothai, A. F. Var¨®n, and Ch. Wunderlich, Nat.Commun.\textbf{5}, 4679 (2014).
\bibitem{35} D. J. Wineland, C. Monroe, W. M. Itano, D. Leibfried, B. E. King, and D. M. Meekhof, J. Res. Natl. Inst. Stand. Technol. \textbf{103}, 259 (1998).
\bibitem{36} D. Kielpinski, C. Monroe, and D. J. Wineland, Nature (London) \textbf{417}, 709 (2002).
\bibitem{37} D. Gottesman and I. L. Chuang, Nature (London) \textbf{402}, 390 (1999).
\bibitem{38} J. D. Jost, J. P. Home, J. M. Amini, D. Hanneke, R. Ozeri, C. Langer, J. J. Bollinger, D. Leibfried, and D. J. Wineland, Nature (London) \textbf{459}, 683 (2009).
\bibitem{39} J. P. Home, D. Hanneke, J. D. Jost, J. M. Amini, D. Leibfried, and D. J. Wineland, Science \textbf{325}, 1227 (2009).
\bibitem{40} D. Hanneke, J. P. Home, J. D. Jost, J. M. Amini, D. Leibfried, and D. J. Wineland, Nat. Phys.\textbf{6}, 13 (2009).

\end{thebibliography}
\end{document}